\documentclass[11pt]{article}

\usepackage{amsmath}
\usepackage{amssymb}
\usepackage{graphicx}
\usepackage{geometry}
\usepackage[normalem]{ulem}
\usepackage{caption}
\usepackage[labelfont=sf]{subcaption}

\usepackage{algorithm}
\usepackage{algorithmic}

\usepackage{hyperref}
\usepackage{enumitem}
\usepackage{xcolor}

\usepackage{natbib}
 \bibpunct[, ]{(}{)}{,}{a}{}{,}%
\newcounter{subsubsubsection}[subsubsection]

\usepackage{chngcntr}

\begin{document}

\title{Hidden in Plain Sight: Detecting Illicit Massage Businesses from Mobility Data}

\author{Roya Shomali, Nickolas K. Freeman$^*$, Gregory J. Bott, Iman Dayarian, Jason M. Parton\\
Department of Information Systems, Statistics, and Management Science,\\
The University of Alabama, Tuscaloosa, AL 35478\\
$^*$Corresponding author: \texttt{freem028@ua.edu}}

\date{}

\maketitle

\begin{abstract}
\textbf{Problem definition}: Illicit massage businesses (IMBs) masquerade as legitimate massage parlors while facilitating commercial sex and human trafficking. Law enforcement must identify these businesses within a dense population of lawful establishments, but investigative resources are limited and the illicit status of each location is unknown until inspection. Detection methods based on online reviews offer some insight, yet operators can manipulate these signals, leaving covert establishments undetected. \textbf{Academic/practical relevance}: IMBs constitute one of the largest segments of indoor sex trafficking in the United States, with an estimated 9,000 establishments. Mobility data offers an alternative to online signals, covering establishments that avoid digital visibility entirely. \textbf{Methodology}: We derive features from mobility data spanning temporal visitation patterns, dwell times, visitor catchment areas, and demand stability. Because confirmed labels exist only for establishments identified through advertising platforms, we employ positive-unlabeled learning to address the label asymmetry in ground truth. \textbf{Results}: The model achieves 0.97 AUC and 0.84 Average Precision. Four operational signatures characterize high-risk establishments: demand consistency, evening-concentrated visits, compressed service durations, and locally drawn clientele. The model produces risk scores for each business-week observation. Aggregating to the business level, prioritizing the highest-risk 10\% of massage establishments captures 53\% of known illicit operations, a 5.3-fold improvement over uninformed inspection. \textbf{Managerial implications}: We develop a decision-support system that produces calibrated prioritization scores for law enforcement, enabling investigators to concentrate inspections on the highest-risk venues. The operational signatures may resist strategic manipulation because they reflect actual operations rather than online signals that operators can control.
\end{abstract}


\section{Introduction}\label{sec:introduction}

Illicit massage businesses (IMBs) masquerade as legitimate massage parlors while deriving a portion of their revenue from commercial sex acts \citep{de2022identifying}. IMBs constitute one of the largest segments of the indoor sex trade in the United States and function as nodes in human trafficking networks. The Polaris Project identifies massage businesses as the second most commonly reported venue type for sex trafficking, with an estimated 9,000 establishments generating \$2.5 billion annually \citep{polaris2017typology, polaris2018imb}. Workers in these establishments are often victims rather than voluntary participants \citep{dank2014estimating}. Traffickers recruit women, frequently undocumented immigrants, through promises of legitimate employment. Once recruited, victims face psychological coercion: threats against family members, fear of deportation, and document confiscation~\citep{polaris2018imb}. Because traffickers rely on these control methods rather than physical restraint, victims rarely cooperate with law enforcement. The combination of a legitimate business facade and psychological control makes IMBs one of the most difficult human trafficking networks to dismantle \citep{tobey2024interpretable, de2022identifying}.

Our detection framework identifies massage establishments likely offering commercial sex services. We do not observe and cannot distinguish human trafficking from voluntary commercial sex work at the point of detection. However, prior research documents that IMBs are a primary venue for labor and sex trafficking \citep{polaris2018imb}, making the identification of such establishments a necessary first step in anti-trafficking enforcement. Throughout this paper, ``illicit'' refers to establishments offering commercial sex services, which enforcement agencies target both for regulatory violations and as potential trafficking venues.

Law enforcement must identify IMBs operating behind a facade of legitimacy within a dense population of lawful massage establishments, creating acute challenges in allocating scarce investigative resources \citep{de2020crime}. A metropolitan area may contain a large number of massage businesses, and inspections require significant investigative resources including surveillance, potential undercover operations, and significant casework to build prosecutable cases. Agencies operate under limited budgets with finite personnel trained for vice investigations and cannot inspect every establishment. They must decide which to investigate when the illicit status of each location is unknown \citep{national2022police}.

The described problem is a resource allocation problem under uncertainty where agencies must allocate limited inspection resources across a large population of massage businesses to maximize the number of illicit operations identified (Section~\ref{sec:problem}). The key input to this optimization problem is a risk estimate for each establishment, but this estimate is difficult to obtain, which can lead agencies to default to reactive strategies, i.e., responding to complaints or tips, that approximate random selection.

Existing detection approaches rely on online signals such as reviews and advertisements \citep{li2023detecting, de2022identifying, tobey2024interpretable}, but these signals are biased toward venues represented in specific online ecosystems and vulnerable to operator manipulation. Anonymized cell phone mobility data offers an alternative, capturing when people visit an establishment, how long they stay, and where they travel from. Together, this information constitutes an operational fingerprint that reflects actual business activity rather than how it presents itself \citep{ghose2019mobile, chang2021nature, wang2022using}. We develop a detection system that extracts 28 features from mobility data across six categories capturing temporal patterns, service durations, visitor origins, and demand stability. To our knowledge, this is the most comprehensive application of mobility data to the detection of illicit activity in the massage industry.

A second challenge is \emph{label asymmetry}. Specifically, the confirmed presence of illicit activity is informative, but the absence of confirmation does not guarantee that a business is legitimate. To accommodate this label asymmetry, we adopt positive-unlabeled (PU) learning \citep{elkan2008pu, bekker2020survey}, scoring each observation by how closely its mobility patterns resemble those of confirmed illicit establishments. Our empirical question is whether illicit establishments leave a detectable signature in mobility data, and we investigate this through three research questions:

\begin{enumerate}
    \item[\textbf{RQ1:}] How effectively can mobility data identify illicit operations across massage businesses?
    \item[\textbf{RQ2:}] Which mobility features drive detection, and what operational patterns do they reveal?
    \item[\textbf{RQ3:}] What inspection efficiency does mobility-based targeting provide?
\end{enumerate}

\noindent Our model generates risk scores for each weekly observation of business behavior. The model achieves 0.97 AUC at the business-week observation level, demonstrating that mobility patterns contain substantial signal for identifying illicit operations. Aggregating the business-week risk scores at the business level enables enforcement prioritization under budget constraints. In our experimentation, an agency allocating inspection resources to the highest-risk 10\% of establishments captures 53\% of known illicit operations, representing a 5.3-fold improvement over uninformed inspection. Inspecting features that drive model performance reveals four operational signatures that characterize high-risk establishments: 1) stable customer demand patterns that persist regardless of typical business cycles, 2) visits concentrated during evening hours when discretion is easier, 3) compressed service durations inconsistent with legitimate therapeutic massage, and 4) elevated foot traffic from nearby origins.

This paper makes four contributions:
\begin{enumerate}
    \item \textbf{Mobility data as a detection signal.} We introduce mobility data as a signal for regulatory screening in adversarial settings. Mobility patterns reflect actual customer behavior (visitation timing, dwell times, and customer origins) that is difficult to falsify without disrupting operations. This data innovation enables detection of establishments that avoid online presence or obfuscate their activities on online platforms.

    \item \textbf{Positive-unlabeled learning for incomplete labels in an illicit context.} Illicit status is confirmed only through investigation or advertising platform linkage; the absence of such evidence provides no information about legitimacy. Our framework treats unlabeled establishments as a mixture of hidden positives and true negatives, enabling detection even when ground truth is fundamentally incomplete. This formulation applies broadly to regulatory screening contexts where violations are discovered but compliance is never directly observed.

    \item \textbf{Linking prediction to resource allocation.} The model produces establishment-level risk estimates that serve as inputs to budget-constrained enforcement optimization (Equation~\eqref{eq:objective}), allowing agencies to shift from reactive to proactive enforcement. The framework generalizes to other regulatory contexts including unlicensed cannabis dispensaries, illegal gambling operations, and unlicensed medical clinics.

    \item \textbf{Empirical characterization of illicit operations.} Beyond demonstrating that mobility data contains predictive signal, we identify four operational signatures that characterize how illicit establishments differ from legitimate ones (Section~\ref{subsec:rq2}). These signatures appear inherent to the illicit business model and difficult to conceal without disrupting the revenue-generating activity itself.
\end{enumerate}

The paper proceeds as follows. Section~\ref{sec:problem} formalizes the resource allocation problem and develops the three challenges that shape our approach: signal vulnerability, label asymmetry, and temporal uncertainty. Section~\ref{sec:lit} reviews related work on IMB detection, machine learning for human trafficking, and positive-unlabeled learning. Section~\ref{sec:modeling} presents our complete modeling framework, including data preparation, feature engineering, the PU learning algorithm, and evaluation methodology. Section~\ref{sec:results} reports detection performance and identifies operational signatures that distinguish illicit establishments. Section~\ref{sec:discussion} discusses implications and limitations.

\section{Problem Statement and Main Challenges}\label{sec:problem}

Detecting illicit massage businesses is a resource allocation problem under uncertainty. This section formalizes the decision problem and develops three structural challenges that shape our methodological approach: signal vulnerability, label asymmetry, and temporal uncertainty.

\subsection{Resource Allocation Formulation}\label{subsec:resource_allocation}

We present a concrete example for a simple variant of the described resource allocation problem that assumes a constant benefit of interdicting an illicit massage business. Specifically, let $i$ index a set of massage business $\mathcal{I}$. Define:
\begin{itemize}
    \item $X_i \in \{0,1\}$: binary decision variable indicating whether business $i$ is selected for inspection,
    \item $c_i$: cost of inspecting business $i$ (investigator time, coordination, undercover operations),
    \item $B$: total investigative capacity, and
    \item $\hat{\delta}_i \in [0,1]$: estimated probability that business $i$ is illicit.
\end{itemize}
The agency solves:
\begin{equation}\label{eq:objective}
\max \sum_{i \in \mathcal{I}} \hat{\delta}_i \cdot X_i, \quad \text{s.t.} \quad \sum_{i \in \mathcal{I}} c_i \cdot X_i \leq B, \quad X_i \in \{0,1\}.
\end{equation}

The challenge is that $\hat{\delta}_i$ is unobserved. Our model estimates $\hat{\delta}_i$ from mobility patterns, providing the key input to the resource allocation problem. Agencies can then solve Equation~\eqref{eq:objective} to prioritize inspections under budget constraints. This framework extends naturally to settings with heterogeneous interdiction benefits, geographic coverage requirements, or capacity constraints across jurisdictions.

\subsection{Signal Vulnerability}\label{subsec:signal_vulnerability}

Existing detection approaches rely on online signals. \citet{li2023detecting} and \citet{de2022identifying} use text analysis to classify online reviews and identify massage businesses exhibiting suspicious patterns. \citet{tobey2024interpretable} combine review text with geographic features to build interpretable risk scores. These methods yield useful information but share a common limitation. They are biased toward venues advertised in specific online ecosystems and cannot capture businesses that avoid online visibility. Operators can curate reviews, modify website content, and alter advertising behavior in response to enforcement attention. A detection system built on signals that operators control is vulnerable to evasion.

Anonymized cell phone mobility data offers an alternative signal source. Mobility data refers to anonymized, aggregated cell phone records of device location derived from GPS signals and mobile applications. Commercial data providers compile these records into visit-level observations indicating when devices appear at specific points of interest, enabling analysis of foot traffic patterns, visitor origins, and temporal visitation behavior without identifying individual users. Such data have enabled research in consumer behavior and retail analytics \citep{ghose2019mobile, athey2018estimating}, epidemiology \citep{chang2021nature}, urban planning, and public policy \citep{wang2022using}. The appeal of mobility data for detection lies in its resistance to manipulation. Online signals are under the operator's control: reviews can be curated, website content modified, and advertising behavior adjusted in response to enforcement attention. Mobility patterns are different. Running a business requires serving customers, and serving customers generates observable patterns of movement and activity. An illicit massage business may cultivate the same online presence as a legitimate spa, but as we will demonstrate, its customer traffic will differ if the services offered differ.

We develop a detection system that exploits this divergence. The system extracts 28 features from mobility data, organized across six categories that capture distinct dimensions of business operations. Temporal features characterize when visits occur and how activity distributes across hours and days. Visit distribution features measure how concentrated visits are across time periods. Market reach features measure how far visitors travel and the geographic spread of the customer base. Service duration features capture dwell time distributions that reflect the nature of services provided. Operational consistency features quantify consistency of customer flow and demand predictability across time. Location context features situate each establishment within its local competitive environment. Together, these features translate raw visitation records into a multidimensional portrait of how each establishment actually operates.

\subsection{Label Asymmetry}\label{subsec:label_asymmetry}

Mobility data addresses signal manipulation, but a second challenge remains: the confirmed presence of illicit activity is informative, but the absence of confirmation is not. We refer to this structural property as \emph{label asymmetry}. Advertising on an adult service platform or appearing on an online review website for illicit massage parlors confirms illicit status. But the absence of such advertising confirms nothing. A massage business without platform presence may be a legitimate therapeutic practice, or it may be an illicit operation that avoids online advertising out of caution. \citet{li2023detecting} identified businesses with illicit indicators on Yelp that never appeared on Rubmaps. Prior work has handled this asymmetry by treating platform presence as a binary label: \citet{diaz2020natural} used Rubmaps listings to label Yelp reviews, assuming all reviews from Rubmaps-listed businesses are illicit and all reviews from unlisted businesses are legitimate. This conflates the absence of evidence with evidence of absence. When we train a classifier on data where unknown cases are labeled legitimate, the algorithm learns to replicate this assumption and will systematically miss illicit establishments that avoid advertising.

We adopt positive-unlabeled (PU) learning to match our method to the structure of the problem \citep{elkan2008pu, bekker2020survey}. PU learning is designed for settings where we observe confirmed positives and an unlabeled pool that contains an unknown mixture of positives and negatives. This matches our setting. Advertising confirms illicit status, but no comparable signal confirms legitimacy. PU learning does not require us to fabricate labels we do not possess. The algorithm learns what distinguishes confirmed illicit behavior from the unlabeled population, scoring each observation by how closely it resembles known illicit patterns. Unlabeled observations that share the mobility signatures of confirmed illicit establishments receive higher risk scores. The model surfaces hidden positives without requiring us to assert which unlabeled observations are legitimate.

\subsection{Temporal Uncertainty}\label{subsec:temporal_uncertainty}

Even confirmed labels carry temporal uncertainty. An establishment may operate legitimately for months before transitioning to illicit services, or may cease illicit activity following a raid. We address this by adopting the business-week as our unit of analysis. An establishment is labeled illicit only during weeks in which it actively advertises on an adult service platform. This ensures that every positive label reflects observed illicit activity in that specific period rather than an assumption that past behavior persists indefinitely. The classifier produces business-week level risk scores that aggregate to the business level for enforcement prioritization.

Together, a mobility-based model that resists signal manipulation and a labeling approach that respects label asymmetry provide the inputs for Equation~\eqref{eq:objective}. The following sections develop the related literature, our modeling framework, and the empirical evaluation.

\section{Related Work}\label{sec:lit}
Our research is related to three streams of prior work: 1) research that performs spatial analysis related to IMBs, 2) research that applies machine learning techniques to human trafficking, and 3) the PU learning literature that motivates our labeling approach. We discuss each in turn.

\subsection{Spatial Analysis of IMBs}
Illicit massage businesses operate at the intersection of commercial sex and labor trafficking, making them critical targets for both criminological research and law enforcement intervention. Their covert nature, often camouflaged as legitimate massage establishments, complicates detection, regulation, and intervention efforts. A growing body of research has examined the geographic distribution of IMBs, with findings organized around three categories of factors: demand-side drivers, supply-side drivers, and locational characteristics.

\paragraph{Demand-side factors.} Several studies find that IMBs concentrate in areas with higher client demand rather than in disadvantaged neighborhoods. \citet{vries_2022_cd} examine IMB distribution across three U.S. states using criminological theories and find that IMBs are located in areas with high residential instability and ethnic diversity but are concentrated in higher-income areas with greater demand rather than disadvantaged neighborhoods. \citet{chin2023and} compare clustering patterns in Los Angeles County and New York City, finding that in NYC, clustering is driven by client demand in high-income business districts. Market forces, rather than local regulations, appear to play a predominant role in shaping locational strategies.

\paragraph{Supply-side factors.} Labor supply also influences IMB location. \citet{chin2023and} find that in Los Angeles County, clustering is primarily influenced by labor supply in immigrant enclaves, contrasting with the demand-driven pattern in NYC. \citet{crotty2018red} examine Houston, Texas from a retail geography perspective and find IMBs locate in neighborhoods with higher percentages of Asian residents and more non-family households, suggesting labor supply considerations. Employment density also correlates with IMB clustering.

\paragraph{Geographic and locational characteristics.} Broader geographic factors also predict IMB presence. \citet{white_2024} employ a nationwide modeling approach and find that state-level regulatory differences, proximity to international airports, and lower rent levels are associated with IMB likelihood. Land use considerations, including urban density and commercial zoning, also shape where these establishments locate \citep{crotty2018red}. While these geographic correlates help explain aggregate patterns, they provide insufficient precision for individual business screening: many legitimate massage establishments share the same locational characteristics as illicit ones. Our approach builds on this geographic foundation by introducing temporal and operational patterns observable at the individual establishment level.

\subsection{Machine Learning and Human Trafficking Detection}
The challenge of distinguishing illicit massage operations from legitimate ones has led researchers to harness data analytics and machine learning. Two primary approaches have emerged: 1) network analysis methods that exploit relational structure in advertising data, and 2) text-based classification methods that identify linguistic patterns associated with trafficking.

Network analysis methods uncover groups involved in sex trafficking by analyzing connections between advertisements. \citet{keskin2021cracking} focus on identifying patterns in sex ads by grouping them based on text, phone numbers, and image hashes. The study then uses these groupings to predict future locations of these ads using different methods, like frequency-based analysis and network simulation. \citet{kjellgren2025geographies} apply social network analysis and principal component analysis (PCA) to analyze the complexity and geographical distribution of online networks in the off-street sex market. \citet{kejriwal2020network} construct ``activity networks'' using sex ad data by linking ad-posting accounts through shared phone numbers and IP addresses. Note that their ability to utilize IP addresses is atypical and is only possible due to a partnership that allowed them to access server-side data from the now-defunct site Backpage.com. \citet{kejriwal2019network} employ network science to assess the quality of information extraction systems in the human trafficking domain without relying on ground-truth data. Their methodology constructs an Attribute Extraction Network from attributes such as names, phone numbers, and addresses from online sex advertisements and analyzes network metrics to evaluate information extraction performance.

Several studies have leveraged text analysis and machine learning techniques to detect human trafficking. Early research relied on predefined keyword lists to detect trafficking-related content. \citet{dubrawski2015leveraging} extract structured features from ad texts, such as keywords and phone numbers, and use these features to train classifiers that distinguish between trafficking-related and non-trafficking ads. Other studies have adopted automated approaches that do not rely on predefined trafficking-related keywords. \citet{zhu2019identification} train a language model on a large dataset of adult service ads to capture linguistic structures specific to trafficking-related content. They then apply machine learning techniques to select key features and classify ads as being related to trafficking.
To automatically learn linguistic patterns associated with trafficking, \citet{esfahani2019context} integrate multiple text feature sets, including topic modeling, average word vectors, and the pre-trained language model BERT for automatically detecting human trafficking advertisements.
\citet{summers2023multi} utilize deep learning models, incorporating text analysis, image/emoji classification, and feature classification, to predict the probability of an ad being associated with sex trafficking.
\citet{li2023detecting} develop a classification framework that leverages both lexicon-based and deep learning representations from BERT and Doc2Vec to detect IMBs from Yelp reviews. \citet{tobey2024interpretable} propose a multi-source machine learning framework to detect illicit massage businesses linked to human trafficking by combining data from data sources such as customer reviews, GIS data, licensing records, and census statistics. While these methods effectively identify suspicious establishments with online visibility, they cannot detect operations that deliberately minimize their online presence. Mobility data complements these approaches by providing signals for establishments that avoid advertising platforms entirely.

\subsection{Positive-Unlabeled Learning}
Positive-unlabeled (PU) learning addresses classification problems where only positive examples and unlabeled data are available, with no confirmed negative instances. This setting arises naturally in domains where positive cases are identified through observation or investigation, but the absence of such identification provides no information about true class membership. \citet{elkan2008pu} established the foundational framework, showing that a classifier trained on positive and unlabeled data can recover the true class-conditional probabilities under the assumption that labeled positives are selected completely at random from all positives. \citet{bekker2020survey} provide a thorough survey of PU learning methods, categorizing approaches into two-step techniques that first identify reliable negatives and single-training-step methods that directly modify the learning objective.

Among the most effective approaches for PU learning is the bagging framework introduced by \citet{mordelet2014bagging}. Rather than attempting to identify reliable negatives, this method treats the unlabeled set as a source of pseudo-negatives. Multiple bootstrap samples are drawn from the unlabeled pool, and an ensemble of classifiers is trained, each distinguishing the positive set from one pseudo-negative sample. Aggregating predictions across the ensemble separates true negatives (which consistently receive low scores) from hidden positives (which resemble the labeled positive set and receive high scores). This approach is robust to contamination of the unlabeled set by hidden positives and does not require estimating the class prior.

PU learning has found applications across domains where confirmed negatives are unavailable. In disease surveillance, cases are identified through diagnosis, while healthy individuals remain unlabeled \citep{yang2012positive}. In fraud detection, confirmed fraud cases serve as positives, while the vast majority of transactions are unlabeled \citep{zhang2021hoba}. In text classification, documents matching specific criteria are labeled positive, while others remain unlabeled \citep{liu2002partially}. Despite this breadth of application, PU learning has not been applied to spatial regulatory screening problems in which illicit facilities are concealed among legitimate establishments.

Our work bridges these streams by introducing mobility data as a manipulation-resistant alternative to online signals and developing a PU learning framework that accommodates the label asymmetry inherent in this domain. While prior spatial analyses have identified geographic correlates of IMB presence, they do not provide operational tools for prioritizing inspections. Machine learning approaches based on online reviews offer detection capabilities but rely on signals that operators can control. We combine a novel data source with a learning framework matched to the label structure, linking prediction to the resource allocation problem that enforcement agencies face. Based on these research streams, we design a three-stage approach. First, we assemble mobility data for massage businesses across the continental United States and identify establishments linked to adult services websites as confirmed illicit. Second, we engineer 28 features capturing temporal, spatial, and operational patterns from weekly mobility records. Third, we train a positive-unlabeled learning classifier that accommodates label asymmetry. The following section details this complete pipeline.

\section{Modeling Framework}\label{sec:modeling}

This section presents our complete framework for detecting illicit massage businesses from mobility data. We begin with a process overview that outlines the pipeline from raw data to establishment-level risk scores. Subsequent subsections detail each component: data preparation, feature engineering, the positive-unlabeled learning algorithm, evaluation methodology, and implementation. Table~\ref{tab:ec-terminology} defines key technical terms used throughout the paper.

\begin{table}[h]
	\centering
	\caption{Key Terminology}
	\label{tab:ec-terminology}
	\begin{tabular}{lp{9.5cm}}
		\textbf{Term} & \textbf{Definition}\\
		\hline
		POI (Point of Interest) & A specific physical business location tracked by the mobility data provider, identified by a unique placekey identifier\\
		CBG (Census Block Group) & A small geographic area defined by the U.S. Census Bureau, typically containing 600-3,000 residents, used for aggregating visitor home locations\\
		NAICS Code & North American Industry Classification System code that categorizes businesses by industry type\\
		Dwell Time & The duration a visitor spends at an establishment during a single visit\\
		PU Learning & Positive-unlabeled learning, a machine learning approach for classification when only positive examples and unlabeled data are available, with no confirmed negatives\\
		AUC & Area Under the ROC Curve, measuring the probability that a randomly selected positive instance ranks higher than a randomly selected negative instance; ranges from 0.5 (random) to 1.0 (perfect)\\
		Average Precision & A metric emphasizing early precision in ranked lists, measuring how efficiently top-ranked items contain true positives\\
		Entropy & An information-theoretic measure of variability; high entropy indicates even distribution across categories, low entropy indicates concentration\\
		Coefficient of Variation & The ratio of standard deviation to mean, expressing relative variability; lower values indicate more stable patterns\\
		Burstiness & A measure of demand irregularity capturing the tendency for visits to cluster in spikes rather than spread evenly; formally, the standard deviation of week-over-week changes normalized by mean visits\\
		\hline
	\end{tabular}
\end{table}

\subsection{Process Overview}
\begin{figure}[htbp]
\centering
\includegraphics[width=1\textwidth]{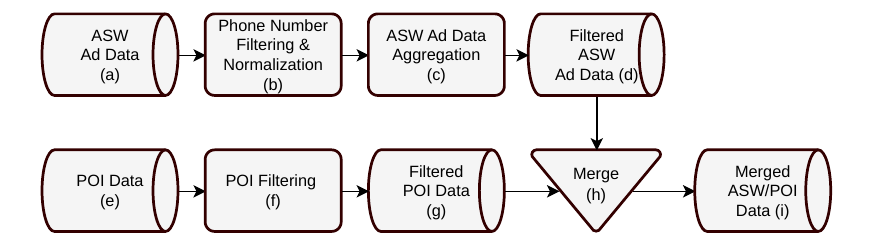}
\caption{Data Merging\label{fig:data_merging}}
\end{figure}

Figure~\ref{fig:data_merging} provides an overview of the data integration workflow. The workflow consists of two parallel data processing pipelines that are merged to create our training data.

The upper pipeline starts with \textit{Adult Services Website (ASW) Ad Data} (a). The phone numbers in this data are normalized and the data is filtered to drop null phone number values (b). Next, the data is aggregated to obtain weekly ad volumes per phone number (c), and this filtered data is persisted (d). The lower pipeline filters \textit{POI Data} (e) to NAICS code 812199, i.e., \textit{Other Personal Care Services}, and retains only establishments whose names contain ``massage'' or ``spa'' (f). This data is persisted (g) and subsequently merged with the filtered ASW data on phone numbers (h). We use a \textit{LEFT} join that retains all filtered POI records and appends matched ASW data (i).

\begin{figure}[htbp]
\centering
\includegraphics[width=0.8\textwidth]{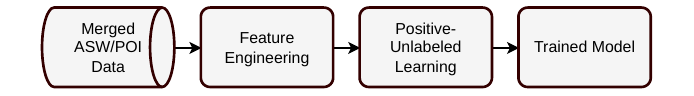}
\caption{Training Pipeline\label{fig:training}}
\end{figure}

The merged dataset produces three observation categories: \textit{Illicit Active} (weeks with ASW ads), \textit{Illicit Quiet} (ASW-linked establishments during weeks without ads), and \textit{Never-ASW} (establishments never linked to ads). An illustrative example of the merged data structure appears in Figure~\ref{fig:ec-merged-data} of Appendix~\ref{ec:merged_data}. Establishments without ASW linkage cannot be assumed non-illicit; the absence of evidence is not evidence of absence. This observation guides our approach to training a classifier, which is depicted in Figure~\ref{fig:training}.

The training pipeline transforms each POI-week record into numerical features capturing temporal, spatial, and operational patterns. These features are passed to a positive-unlabeled (PU) learning algorithm that treats ASW-linked observations as confirmed positives and all other observations as unlabeled (rather than negative), producing a trained model that scores each POI-week by its likelihood of illicit activity.

\subsection{Data}\label{sec:data}
This subsection describes the data sources, labeling strategy, and dataset preparation for our analysis. We use mobility data from a commercial provider to characterize establishment behavior and link advertising records from adult services websites to identify confirmed illicit operations.

\subsubsection{Mobility Data.}\label{subsec:feature_data}
We obtain the feature data for this study from \href{https://advanresearch.com/}{Advan} via a subscription to the \href{https://www.deweydata.io/}{Dewey Data platform}. Specifically, we use a data product called the \textit{Weekly Patterns Dataset}. This dataset includes visitor and demographic aggregations for US POIs over a week, with individual rows corresponding to data for a single POI from Monday to the end of the day on Sunday each week. Each record includes data for nearly fifty variables. Table~\ref{tab:mobility-data-variables} describes the key variables we use.

\begin{table}[htbp]
\centering
\caption{Mobility Data Variables}
\label{tab:mobility-data-variables}
\begin{tabular}{lp{9.5cm}}
\hline
Column Name & Description\\
\hline
\texttt{placekey} & A unique and persistent ID tied to a Point of Interest (POI)\\
\texttt{naics\_code} & The NAICS (North American Industry Classification System) code value that represents the industry associated with the POI\\
\texttt{city} & The city where the POI is located\\
\texttt{region} & The state, province, county, or equivalent of how ``region'' is understood in the country in which the POI is located\\
\texttt{postal\_code} & The postal code of the POI\\
\texttt{open\_hours} & A JSON string with days as keys and opening \& closing times (in the POI's local time) as values\\
\texttt{date\_range\_start} & The start date of the data collection period\\
\texttt{date\_range\_end} & The end date of the data collection period\\
\texttt{raw\_visit\_counts} & The number of unique visits to the POI during the date range\\
\texttt{raw\_visitor\_counts} & The number of unique visitors to the POI during the date range\\
\texttt{visits\_by\_each\_hour} & The number of visits to the POI, broken down by each hour of the week (168 values)\\
\texttt{poi\_cbg} & The Census Block Group (CBG) of the POI\\
\texttt{visitor\_home\_cbgs} & The number of visitors to the POI from each CBG based on the visitor's home location\\
\texttt{visitor\_daytime\_cbgs} & The number of visitors to the POI from each CBG based on the visitor's primary daytime location between 9 a.m.--5 p.m.\\
\texttt{bucketed\_dwell\_times} & The distribution of visit dwell times based on pre-specified buckets. The key is the dwell time range in minutes, and the value is the number of visits within that range\\
\hline
\end{tabular}
\end{table}

Two aspects of the data are worth noting. First, Advan computes all POI metrics using the POI's geometry without applying dwell-time filters. One justification provided by Advan for this decision is that dwell-time filtering reduces correlation with ground-truth revenue and transaction counts across 1,500 publicly traded tickers. Second, home and work locations are assigned based on historical device patterns (most frequented nighttime and daytime buildings) over the full observation period, and are not updated weekly.

\subsubsection{Ground Truth Labels.}\label{subsec:label_data}

We obtain ground truth labels from an adult services website (ASW) that functions as an online classified advertising platform for escort and adult services. Service providers post advertisements containing contact information, service descriptions, and location details. We link ASW ads to physical massage establishments by matching phone numbers. Phone numbers provide the most direct link between online advertisements and physical business locations, and this matching approach is consistent with prior work in this domain \citep{keskin2021cracking, kejriwal2020network}. When a phone number appearing in ASW advertisements matches the contact number for a massage business in the Advan dataset, we identify that establishment as engaging in illicit activity. This matching strategy captures massage parlors that use the ASW to solicit clients for commercial sex services. We construct labels at the POI-week level rather than the establishment level. Weeks containing active ASW advertisements represent confirmed periods of illicit operation. This temporal alignment between advertising activity and mobility patterns produces three mutually exclusive observation categories:

\begin{itemize}
    \item \textbf{Illicit Active}: POI-week observations in which the establishment posted at least one ASW advertisement during that week ($N = 18,903$). These observations represent weeks of confirmed illicit operation and serve as positive instances in our classification framework.

    \item \textbf{Illicit Quiet}: POI-week observations for ASW-matched establishments during weeks without active advertising ($N = 23,234$). While these establishments have demonstrated illicit activity at other times, we cannot confirm active illicit operation during these specific weeks.

    \item \textbf{Never-ASW}: POI-week observations for establishments with no ASW presence throughout our study period ($N = 849,639$). These observations constitute the unlabeled set and may represent either legitimate businesses or illicit operations that utilize alternative advertising channels.
\end{itemize}

\noindent Our training process uses only Illicit Active and Never-ASW observations for training, isolating the clearest behavioral signal.

\subsubsection{Final Dataset.}\label{subsec:dataset_prep_for_modeling}
As described in the POI Filtering step of Figure~\ref{fig:data_merging}, the process begins with the weekly patterns data. We apply five separate filters to this data. The first filter limits the data to include records corresponding to businesses associated with the NAICS code 812199, corresponding to ``other personal care services.'' This NAICS classification includes massage parlors and other services, such as day spas and tattoo parlors. Second, we require establishment names to contain ``massage'' or ``spa'' to focus on the target population. Third, we filter the data only to include businesses in the contiguous US. Fourth, we filter the data only to include observations starting January 1, 2024. Fifth, there are cases where the data for a business may stop due to the business closing. We limit our analysis to establishments with continuous operation since the specified start date by looking for null values in the \texttt{visits\_by\_each\_hour} field. These filters yield 891,776 POI-week observations across 19,567 unique establishments. Of these, we can match 920 establishments to ASW ads, contributing 42,137 POI-weeks (18,903 during active advertising and 23,234 during \textit{quiet} periods). The remaining 18,647 establishments contribute 849,639 POI-weeks to the unlabeled pool observed from January 2024 through December 2024.

\begin{figure}[htbp]
\centering
\includegraphics[width=\textwidth]{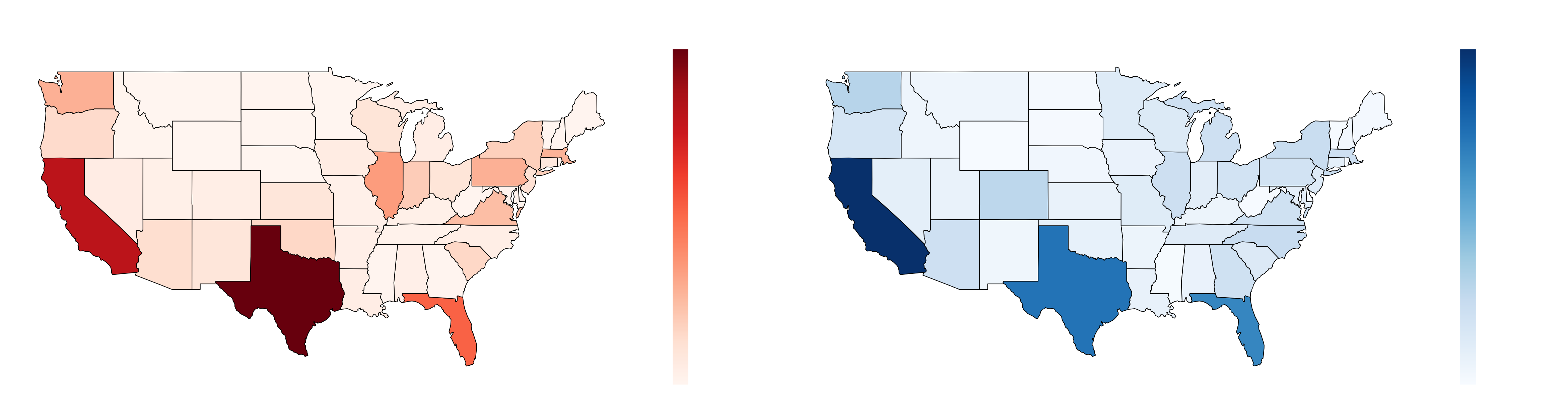}
\caption{Geographic Distribution of POI-Weeks\label{fig:geographic_coverage}}
\end{figure}

Figure~\ref{fig:geographic_coverage} displays the geographic distribution
of POI-week observations. The left subplot shows observations from establishments with ASW ad observations and the right subplot shows all observations in the dataset. As the figure shows, the spatial distribution of establishments with ASW ad observations closely follows the overall distribution, which suggests that the labeled data provides geographic coverage across regions. This geographic alignment reduces concerns that the model learns region-specific artifacts rather than generalizable operational patterns.

\subsection{Feature Engineering}\label{subsec:features}

We engineer 28 features from mobility data across the six categories introduced in Section~\ref{subsec:signal_vulnerability}, plus visit volume as a control variable. Table~\ref{tab:feature-summary} summarizes these features; detailed formulas and computational specifications appear in Appendix~\ref{app:features}.

\begin{table}[htbp]
\centering
\caption{Summary of Engineered Features}
\label{tab:feature-summary}
\begin{tabular}{llp{7cm}}
\hline
\textbf{Category} & \textbf{\# Features} & \textbf{Description} \\
\hline
Temporal patterns & 8 & Proportion of visits in six 4-hour windows (early morning through late night), weekend share, Friday--Saturday share \\
Visit distribution & 3 & Hourly entropy, daily entropy, peak hour ratio \\
Service duration & 3 & Share of short ($<$5 min), medium (5--60 min), and long ($>$60 min) visits \\
Market reach & 6 & Share of visitors from six distance bands (0--1 mi through $>$60 mi) \\
Operational consistency & 5 & Coefficient of variation, trend, burstiness, maximum jump, active ratio \\
Location context & 2 & Log CBG area, county partisan index \\
Volume control & 1 & Log weekly visits \\
\hline
\end{tabular}
\end{table}

\textit{Temporal patterns} capture when customers visit. We partition the 168-hour week into six non-overlapping 4-hour windows and compute the proportion of visits in each, plus weekend and Friday--Saturday shares. These features distinguish establishments with standard business-hour operations from those serving evening or late-night clientele.

\textit{Visit distribution} features measure how concentrated visits are across time. Hourly and daily entropy (see Appendix~\ref{app:features}) quantify whether visits spread evenly or cluster into narrow windows; lower entropy indicates concentration in specific time periods. Peak hour ratio identifies establishments with pronounced rush periods.

\textit{Service duration} features characterize how long customers stay. The data provider (Advan) reports dwell times in pre-specified buckets ($<$5, 5--10, 11--20, 21--60, 61--120, 121--240, and $>$240 minutes). We aggregate these into three categories: short ($<$5 min), medium (5--60 min), and long ($>$60 min). These categories capture the distinction between brief visits, standard service durations, and extended sessions. Legitimate massage sessions typically last 30--90 minutes \citep{field2016massage}; different duration profiles may indicate different service models.

\textit{Market reach} features describe the geographic origin of visitors. We compute great-circle distances (shortest path across Earth's surface) from visitor home CBG centroids to each POI using the Haversine formula and bin into six categories from very local ($<$1 mile) to out-of-area ($>$60 miles). Local versus regional draw may reflect different business models and marketing channels. Elevated local market reach may also reflect word-of-mouth referral channels, which is itself a distinguishing operational signature.

\textit{Operational consistency} features measure traffic stability across weeks. Coefficient of variation, burstiness, and maximum jump capture demand volatility. Trend identifies growing or declining establishments. Active ratio flags periods of apparent inactivity. Because these features are computed across multiple weeks, we recompute them within each evaluation fold using only training-period data, ensuring that held-out observations do not influence feature values during evaluation.

\textit{Location context} features allow the model to calibrate behavioral signals across geographic settings. Log-transformed CBG area serves as a proxy for urban density. We use county-level partisan index scores to serve as a proxy for regulatory environment variation across jurisdictions. Prior research suggests enforcement intensity and resource allocation for vice crimes correlate with local political composition \citep{white_2024}. We include this control to account for geographic variation rather than as a predictive feature of an illicit operation.

\subsection{Positive-Unlabeled Learning}\label{sec:pu_learning}

\paragraph{Intuition.} The label asymmetry described in Section~\ref{subsec:label_asymmetry} means that standard classification, which treats all unknowns as negatives, would learn to distinguish ``those who advertise online'' from ``everyone else'' rather than ``illicit'' from ``legitimate.'' Hidden illicit establishments that persist in the negative class bias the decision boundary.

Positive-unlabeled (PU) learning addresses this by treating the unknown set as a \textit{mixture} of hidden positives and true negatives. The approach we adopt trains an ensemble of classifiers, each using a different random sample of unknowns as surrogate negatives. Averaging across the ensemble separates true negatives (which consistently receive low scores) from hidden positives (which share behavioral patterns with confirmed positives and consistently receive high scores). Building on foundational work in PU learning \citep{elkan2008pu, bekker2020survey}, we combine the \textit{Spy technique} for reliable negative identification \citep{liu2002partially} with bootstrap aggregation over pseudo-negative samples \citep{mordelet2014bagging}.

\subsubsection{PU Learning Problem Setup.}

Let $x_{it} \in \mathbb{R}^d$ denote the feature vector for establishment $i$ in week $t$, which we construct from mobility data. Let $y_{it} \in \{0,1\}$ denote the true (latent) operational status, where $y_{it} = 1$ indicates illicit activity during week $t$. We do not observe $y_{it}$ directly; instead, we observe advertising behavior. Let $a_{it} = 1$ if
establishment $i$ posts an ASW advertisement during week $t$. Our identifying assumption is:
\begin{equation}
    a_{it} = 1 \implies y_{it} = 1,
\end{equation}
i.e., advertising on an ASW constitutes sufficient evidence of illicit operation. The converse does not hold, i.e., $a_{it} = 0$ provides no information about $y_{it}$. This assumption partitions observations into two sets:
\begin{itemize}
    \item \textbf{Positive set} $\mathcal{P}$: POI-weeks with
    $a_{it} = 1$ (Illicit Active observations).
    These are confirmed illicit.
    \item \textbf{Unlabeled set} $\mathcal{U}$: POI-weeks with
    $a_{it} = 0$ for establishments never appearing on the ASW
    (Never-ASW observations). These may be
    legitimate or illicit through other channels.
\end{itemize}
We exclude Illicit Quiet observations (weeks without advertising
for ASW-matched establishments) from both sets. Including them
as positives dilutes signal quality; including them as unlabeled
risks contaminating pseudo-negative samples with known illicit
establishments. Section~\ref{subsec:rq1} validates
this design choice.

\paragraph{Learning Objective.}
Enforcement agencies solve the budget-constrained targeting
problem~\eqref{eq:objective}, which requires risk estimates
$\hat{\delta}_i$ for each establishment. Our task is to learn
a scoring function $\hat{r}: \mathbb{R}^d \rightarrow [0,1]$ from
$\mathcal{P}$ and $\mathcal{U}$ such that $\hat{r}(x_{it})$ reflects
the likelihood of illicit operation. This relies on hidden
positives in $\mathcal{U}$ sharing behavioral signatures with
confirmed positives in $\mathcal{P}$. Weekly scores aggregate
to establishment-level estimates:
\begin{equation}\label{eq:aggregation}
    \hat{\delta}_i = g\bigl(\{\hat{r}(x_{it})\}_{t}\bigr)\quad \forall i \in \mathcal{I}.
\end{equation}
Section~\ref{subsec:RQ3} specifies the aggregation function $g$.

Treating $\mathcal{U}$ as confirmed negatives would penalize the classifier for assigning high scores to hidden positives, causing it to distinguish \textit{platform-advertised} from \textit{non-advertised} establishments rather than \textit{illicit} from \textit{legitimate} (see Appendix~\ref{ec:standard_fails} for a formal derivation).

\subsubsection{Learning from Positive and Unlabeled Data.}

The PU Bagging algorithm operationalizes the ensemble intuition from Section~\ref{sec:pu_learning} by repeatedly sampling pseudo-negative sets from $\mathcal{U}$. The key assumption is that while any single instance from $\mathcal{U}$ may be positive or negative, the \textit{majority} of a random subset belongs to the negative class. Averaging across classifiers trained on different subsets provides robustness to any particular grouping of pseudo-negatives.

In each iteration $k$, we sample a pseudo-negative set $\mathcal{N}_k$ of size $|\mathcal{P}|$ from $\mathcal{U}$ without replacement. This balanced sampling ensures the base classifier is not overwhelmed by the larger unlabeled set. The base classifier $h_k$ is trained to distinguish $\mathcal{P}$ from $\mathcal{N}_k$, producing probability estimates $h_k(x) \in [0,1]$.

\begin{algorithm}[htbp]
\caption{PU Bagging}
\label{alg:pu-bagging}
\begin{algorithmic}[1]
\REQUIRE Positive set $\mathcal{P}$, Unlabeled set $\mathcal{U}$, Number of iterations $K$, Base classifier $\mathcal{H}$
\ENSURE Risk scores $\hat{r}(x)$ for all $x \in \mathcal{P} \cup \mathcal{U}$
\STATE Initialize score accumulator: $R(x) \leftarrow 0$ for all $x$
\FOR{$k = 1$ to $K$}
    \STATE Sample $\mathcal{N}_k \subset \mathcal{U}$ uniformly at random, $|\mathcal{N}_k| = |\mathcal{P}|$
    \STATE Construct training set: $\mathcal{T}_k = \{(x, 1) : x \in \mathcal{P}\} \cup \{(x, 0) : x \in \mathcal{N}_k\}$
    \STATE Train classifier: $h_k \leftarrow \mathcal{H}(\mathcal{T}_k)$
    \STATE Update scores: $R(x) \leftarrow R(x) + h_k(x)$ for all $x \in \mathcal{P} \cup \mathcal{U}$
\ENDFOR
\STATE Compute final scores: $\hat{r}(x) \leftarrow R(x) / K$ for all $x$
\RETURN $\hat{r}(x)$
\end{algorithmic}
\end{algorithm}

The final risk score for each instance is the average prediction across all $K$ iterations:

\begin{equation}
    \hat{r}(x) = \frac{1}{K} \sum_{k=1}^{K} h_k(x).
\end{equation}

\noindent Averaging over $K$ classifiers stabilizes score estimates. True negatives receive consistently low scores across iterations, while hidden positives receive consistently high scores, producing separation without confirmed negative examples.

Algorithm~\ref{alg:pu-bagging} follows the standard PU Bagging framework of \citet{mordelet2014bagging}. A remaining challenge is evaluation. With no confirmed negatives, standard holdout metrics such as precision and recall are not directly applicable. We address this through a \textit{spy technique} \citep{liu2002partially} that withholds a fraction of known positives and embeds them in the unlabeled pool, simulating deployment conditions where some illicit establishments remain undetected. The recovery rate of these ``spies'' serves as our primary evaluation measure. Section~\ref{subsec:evaluation} details the full evaluation protocol.

\subsection{Evaluation Framework}\label{subsec:evaluation}

We evaluate our approach at two levels: POI-week predictions and business-level rankings. POI-week evaluation assesses the model's ability to identify illicit activity during specific time periods. Business-level evaluation measures screening efficiency for prioritizing inspections across establishments.

\subsubsection{POI-Week Evaluation.}
\paragraph{The Spy Technique.}

Standard holdout evaluation may be optimistic because held-out positives never experience being treated as pseudo-negatives during training. Inspired by the \textit{spy technique} \citep{liu2002partially}, we hide 20\% of known positives ($\mathcal{S} \subset \mathcal{P}$) in the unlabeled pool during training, simulating deployment conditions where hidden positives are sometimes sampled as pseudo-negatives. The protocol partitions confirmed positives into 80\% for training ($\mathcal{P}_{\text{train}}$) and 20\% as spies, constructs an augmented unlabeled pool $\mathcal{U}' = \mathcal{U} \cup \mathcal{S}$, trains PU Bagging on $\mathcal{P}_{\text{train}}$ and $\mathcal{U}'$, and evaluates how well the model ranks spies above truly unlabeled observations.

\paragraph{Evaluation Metrics.}
We report metrics relevant to operational screening in the PU learning setting. Note that standard precision, recall, and F1 are \textit{not valid} in PU learning because we cannot identify true negatives, i.e., what appears to be a ``false positive'' on unlabeled data may actually be a correctly identified hidden positive. Instead, we use:

\begin{itemize}
	\item \textbf{Discriminative power (AUC):} The probability that a randomly chosen spy observation scores higher than a randomly chosen truly unlabeled observation:
	\begin{equation}
		\text{AUC} = P\bigl(\hat{r}(x_{it}) > \hat{r}(x_{jt'})\bigr)
		\quad \text{for } (i,t) \in \mathcal{S}, \; (j,t') \in \mathcal{U},
	\end{equation}
	where values near 1.0 indicate strong separation.
	\item \textbf{Ranking quality (Average Precision):} AP summarizes the precision-recall curve when treating spy observations as the retrieval target:
	\begin{equation}
		\text{AP} = \sum_{k} (R_k - R_{k-1}) \cdot P_k,
	\end{equation}
	where $P_k$ and $R_k$ are precision and recall at the $k$-th ranked observation. This measures how early spy observations appear in the ranked list without requiring true negative labels. High AP indicates that top-scored observations efficiently surface hidden illicit cases.
	\item \textbf{Recovery rate at threshold $\tau$:} The fraction of hidden positive observations scoring above $\tau$:
	\begin{equation}
		R(\tau) = \frac{|\{(i,t) \in \mathcal{S} : \hat{r}(x_{it}) > \tau\}|}{|\mathcal{S}|}
	\end{equation}
	This directly answers the question ``If we flag all POI-weeks scoring above	$\tau$, what fraction of hidden illicit observations do we recover?'' We evaluate recovery at several thresholds (e.g., $\tau\in\{0.5,0.6,0.7\}$).
\end{itemize}

\subsubsection{Business-Level Screening.}

The PU Bagging procedure produces a risk score $\hat{r}(x_{it})$ for each POI-week observation. For operational deployment, we aggregate to establishment-level estimates $\hat{\delta}_i$ as defined in~\eqref{eq:aggregation}. Let $\mathcal{T}_i$ denote the set of weeks observed for establishment $i$. We consider three aggregation functions $g$:
\begin{align}
    \hat{\delta}_i^{\text{mean}} &= \frac{1}{|\mathcal{T}_i|}
        \sum_{t \in \mathcal{T}_i} \hat{r}(x_{it}) \\
    \hat{\delta}_i^{\text{max}} &= \max_{t \in \mathcal{T}_i}
        \hat{r}(x_{it}) \\
    \hat{\delta}_i^{\text{min}} &= \min_{t \in \mathcal{T}_i}
        \hat{r}(x_{it})
\end{align}
Mean aggregation ($\hat{\delta}_i^{\text{mean}}$) captures the consistent risk level of establishments across all periods. Max aggregation ($\hat{\delta}_i^{\text{max}}$) captures the maximum risk level across all periods. Finally, min aggregation ($\hat{\delta}_i^{\text{min}}$) captures the minimum risk level across all periods. We evaluate the performance of these three methods in Section~\ref{subsec:RQ3}.

\paragraph{Business-Level Evaluation.}
We partition confirmed illicit businesses using 5-fold cross-validation. In each fold, 80\% of illicit establishments
(with all their weeks) are used for training; 20\% are held out. The model never observes any week from holdout establishments during training, simulating deployment conditions where investigators encounter previously unknown establishments.

Under label asymmetry, where illicit status is confirmed only for positive cases, standard precision and recall are
undefined. We measure performance exclusively on known positives. Let $\mathcal{I}^+_{\text{holdout}}$ denote the set of holdout illicit establishments. Define the \textit{coverage} at budget $K$ as the fraction of the holdout illicit establishments appearing in the top $K$\% of ranked output:
\begin{equation}
\text{Coverage}(K) = \frac{|\{i \in \mathcal{I}^+_{\text{holdout}} :
    \hat{\delta}_i \geq \tau_K\}|}{|\mathcal{I}^+_{\text{holdout}}|},
\end{equation}
where $\tau_K$ denotes the $(100-K)$th percentile of $\{\hat{\delta}_i\}_{i \in \mathcal{I}}$.

\subsection{Implementation}\label{subsec:implementation}

We implement the PU learning approach in Python using the \texttt{scikit-learn} library. Preliminary experiments comparing base learners (logistic regression, gradient boosting, and random forest) indicated that Random Forest achieved the highest screening AUC, leading to its selection for the final model. We conduct a grid search over iteration count ($K \in \{20, 30, 50, 75, 100\}$), tree depth ($\{10, 15, 20, 25, 30, 40\}$), and forest size ($\{50, 100, 200\}$). The best configuration uses $K=50$ iterations, each training a Random Forest of 100 trees with maximum depth of 30 and minimum leaf size of 5 observations.

\section{Experimental Results}\label{sec:results}

We evaluate our mobility-based detection framework in light of our three research questions. RQ1 assesses whether mobility data patterns contain sufficient signal to distinguish illicit operations from legitimate massage businesses. RQ2 examines which mobility features drive detection performance and what operational signatures characterize high-risk establishments. RQ3 quantifies the practical efficiency gains from mobility-based targeting compared to uninformed inspection strategies.

\subsection{RQ1: Can Mobility Data Detect Illicit Activity?}\label{subsec:rq1}

Our framework combines engineered mobility features capturing temporal, spatial, and operational patterns with a PU Bagging classifier trained on temporally-aligned ground truth labels. To assess detection performance, we employ the spy technique, which provides conservative estimates by hiding a subset of known positives in the unlabeled pool during training. This evaluation protocol simulates realistic deployment conditions where some illicit establishments remain undetected and potentially bias the training process.

\subsubsection{Detection Performance.}

Table~\ref{tab:main-results} presents detection performance for our primary specification (Approach A), which trains exclusively on active advertising weeks. An AUC of 0.973 indicates that a randomly selected illicit establishment scores higher than a randomly selected unlabeled establishment 97.3\% of the time. The Average Precision of 0.843 complements AUC by emphasizing that the ranked list efficiently surfaces illicit cases. Regarding recovery rates, at a threshold of 0.5, the model recovers 86.5\% of hidden illicit operations despite these test cases being embedded in the unlabeled pool during training. At the more conservative threshold of 0.7, the model still recovers over one-third of illicit cases, specifically those with the strongest behavioral signatures. High-scoring unlabeled locations represent candidate IMBs not captured in advertising data, potentially surfacing previously undetected operations.

\begin{table}[htbp]
\centering
\caption{Detection Performance (Approach A, Spy Technique)}
\label{tab:main-results}
\begin{tabular}{lc}
\hline
\textbf{Metric} & \textbf{Value} \\
\hline
AUC & 0.973 \\
Average Precision & 0.843 \\
\hline
Recovery Rate @ 0.5 & 86.5\% \\
Recovery Rate @ 0.6 & 65.1\% \\
Recovery Rate @ 0.7 & 37.4\% \\
\hline
\end{tabular}
\end{table}

Table~\ref{tab:score-distributions} presents mean risk scores across observation categories. Spies score 0.652 compared to 0.307 for unlabeled (Never-ASW) observations, providing clear separation despite the adversarial setup. Illicit Quiet observations (mean score 0.519) fall between training positives and unlabeled observations, demonstrating that the model captures persistent operational signatures rather than advertising-correlated artifacts. This supports the core premise that illicit mobility patterns persist regardless of advertising activity.

\begin{table}[htbp]
	\centering
	\caption{Mean Risk Scores by Observation Category}
	\label{tab:score-distributions}
	\begin{tabular}{lc}
		\hline
		\textbf{Category} & \textbf{Mean Score} \\
		\hline
		Training Positives (Illicit Active) & 0.775 \\
		Spies (Hidden Illicit Active) & 0.652 \\
		Illicit Quiet & 0.519 \\
		Unlabeled (Never-ASW) & 0.307 \\
		\hline
	\end{tabular}
\end{table}

\begin{figure}[htbp]
\centering
\includegraphics[width=0.6\textwidth]{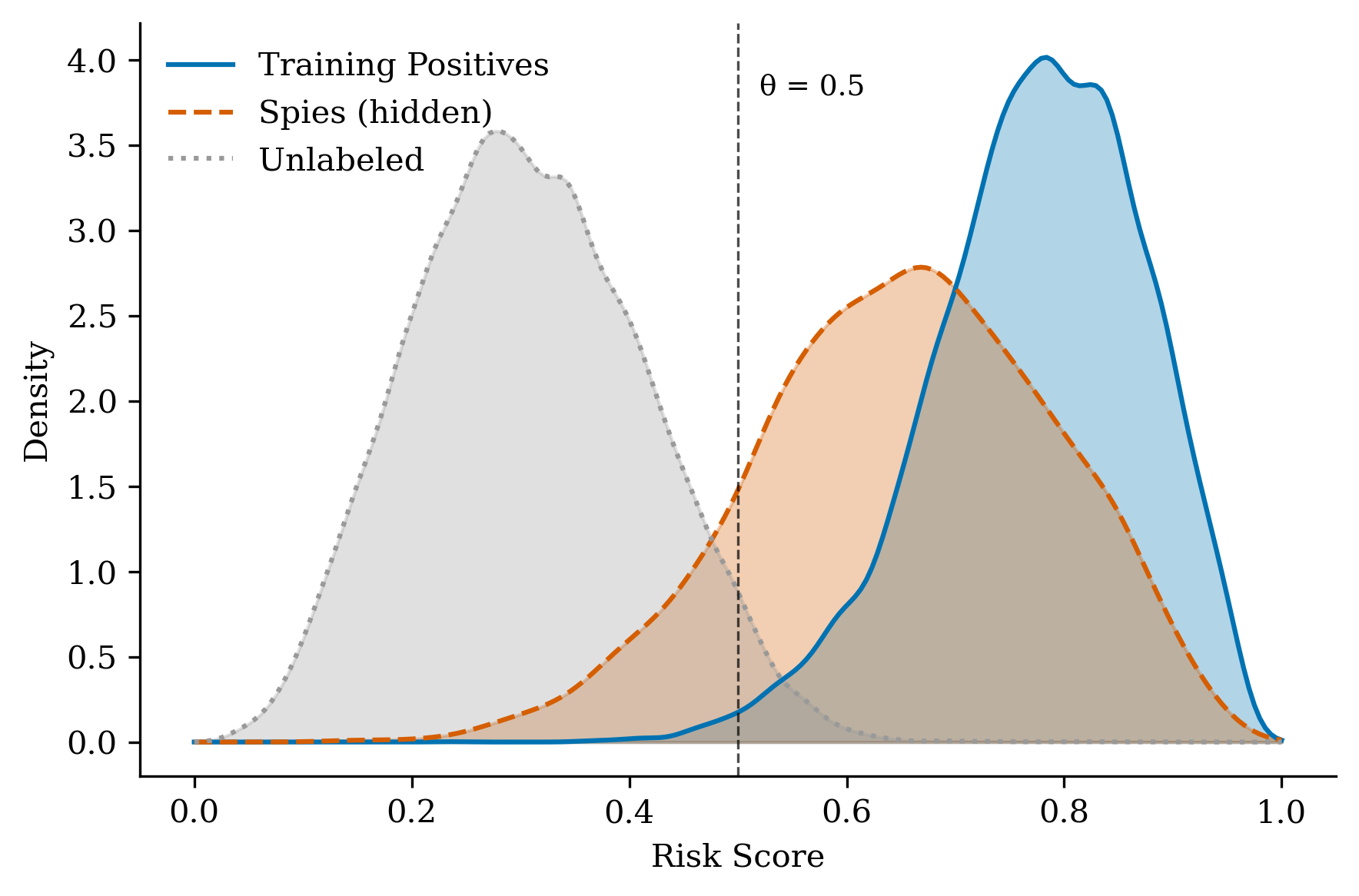}
\caption{Score Distributions by Observation Category\label{fig:score-distributions}}
\end{figure}

Figure~\ref{fig:score-distributions} shows the full score distributions. The separation between groups is clear. Specifically, unlabeled observations concentrate below 0.5, while spies and training positives concentrate above.

\subsubsection{Validation of Temporal Labeling Strategy.}

Our methodology distinguishes between active and quiet weeks for confirmed illicit establishments. Table~\ref{tab:approach-comparison} compares three specifications: 1) Approach A, which includes Illicit Active observations in $\mathcal{P}$ and does not use Illicit Quiet observations, 2) Approach B, which includes Illicit Quiet observations in $\mathcal{U}$, and 3) Approach C, which includes Illicit Quiet observations in $\mathcal{P}$. The results show that Approach A achieves the highest performance. Average Precision drops from 0.843 to 0.659 when quiet weeks are treated as positives (Approach C), reflecting signal dilution from mixing uncertain observations with confirmed active weeks. Approach B shows minor degradation because quiet observations represent a small fraction of the unlabeled pool.

\begin{table}[htbp]
\centering
\caption{Comparison of Approaches}
\label{tab:approach-comparison}
\begin{tabular}{lccc}
\hline
\textbf{Metric} & \textbf{A: Active Only} & \textbf{B: Quiet in Unlabeled} & \textbf{C: Quiet as Positive} \\
\hline
Spy AUC & 0.973 & 0.964 & 0.945 \\
Spy AP & 0.843 & 0.832 & 0.659 \\
Recovery @ 0.5 & 86.5\% & 82.2\% & 82.8\% \\
\hline
\end{tabular}
\end{table}

\subsection{RQ2: Which Mobility Features Drive Detection?}\label{subsec:rq2}

Understanding \textit{why} the model flags specific establishments is important for regulatory deployment. We measure the impact of features using permutation importance. Permutation importance measures the drop in AUC when the values of a feature are randomly shuffled. Large drops in AUC indicate that the shuffling significantly degrades model performance, hence, indicating the feature is important. For this investigation, we compare the top 10\% (high-risk) observations to the bottom 10\% (low-risk) observations. Four operational signatures emerge from this analysis.

\subsubsection{Signature 1: Demand Stability}\label{subsec:sig1}

\begin{figure}[htbp]
\centering
\includegraphics[width=0.35\textwidth]{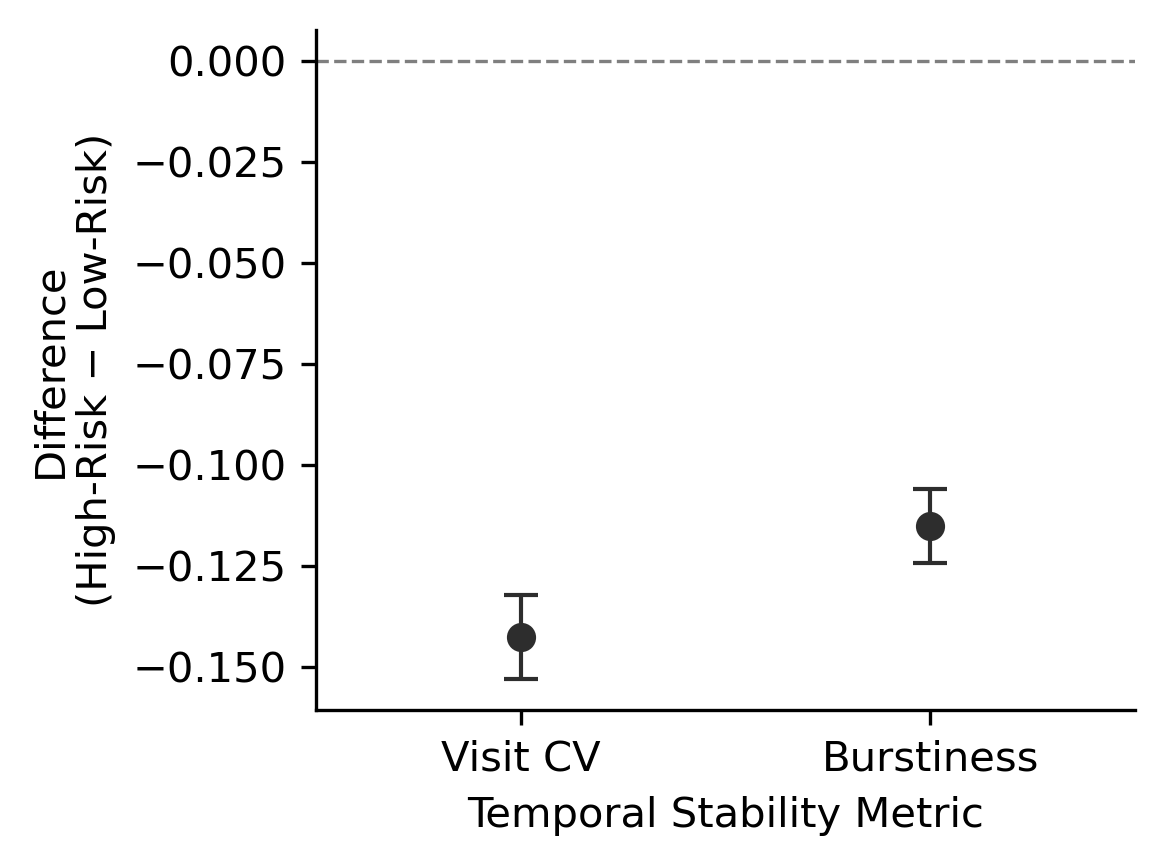}
\caption{Demand Stability Metrics: Difference Between High-Risk and Low-Risk Establishments\label{fig:temporal_stability}}
\end{figure}

The strongest identified predictor of illicit activity is the temporal consistency of customer demand. The temporal consistency features (coefficient of variation, trend, burstiness, maximum jump, and active ratio) collectively account for 36.4\% of model importance. Figure~\ref{fig:temporal_stability} displays the difference in these metrics between high-risk and low-risk establishments. Negative values indicate that high-risk establishments exhibit lower values for that metric.

The figure shows that high-risk establishments exhibit lower coefficient of variation (by $\sim$0.145) and burstiness (by $\sim$0.115), indicating more stable and predictable traffic. These observations are consistent with a loyal base of repeat customers rather than variable demand driven by new clients and seasonal fluctuations. This stability connects to the underlying business model. Illicit operations may cultivate a regular client base through repeat visits and informal referrals, producing demand patterns that are unusually consistent compared to legitimate massage businesses, which experience more variability from new-client acquisition, seasonal trends, and promotional campaigns.

\subsubsection{Signature 2: Time-Shifted Operations}\label{subsec:sig2}

\begin{figure}[htbp]
\centering
\includegraphics[width=0.6\textwidth]{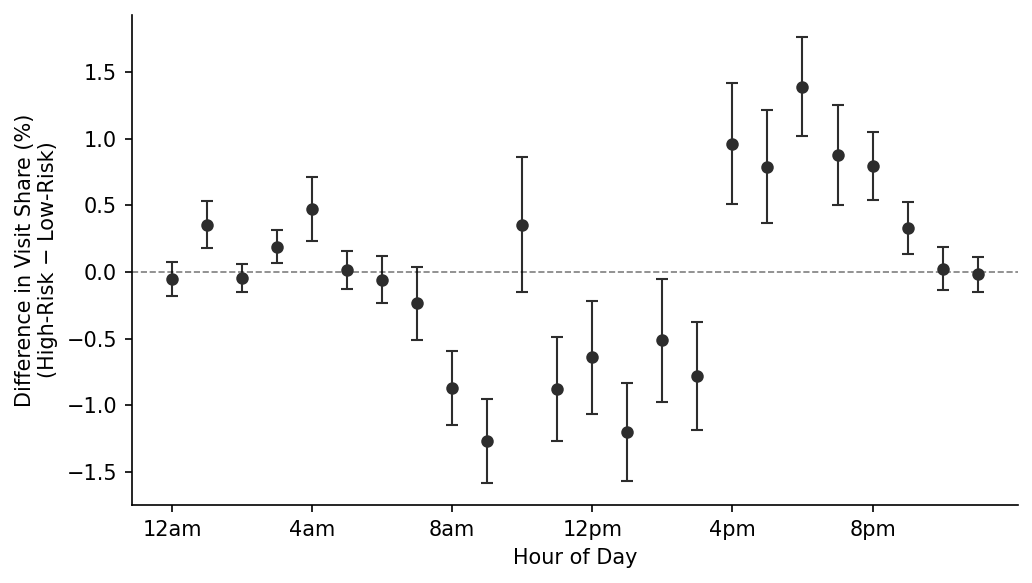}
\caption{Hourly Visit Difference by Risk Category\label{fig:hourly_patterns}}
\end{figure}

Time-of-day features contribute 10.6\% of model importance. Figure~\ref{fig:hourly_patterns} shows that high-risk establishments receive fewer visits during standard business hours (differences most negative around 9am and 1pm, $\sim$-1.3 pp) and more visits in the evening (differences peak near 6pm at $\sim$+1.4 pp, remaining elevated through 10pm). This pattern is consistent with clientele seeking services after work hours. Evening operations also offer operational advantages for illicit businesses. Reduced foot traffic in commercial areas during off-peak hours provides greater discretion, and the target clientele is available after standard work hours. While some legitimate spas offer evening appointments, the \textit{systematic} shift toward evening hours across the full week distinguishes high-risk establishments from those that simply extend their hours occasionally.

\subsubsection{Signature 3: Visit Duration}\label{subsec:sig3}

Dwell time patterns contribute 9.4\% of model importance. Table~\ref{tab:dwell} shows that 75.8\% of high-risk observations correspond to visits lasting less than 5 minutes versus 65.9\% for low-risk, while long visits ($>$60 min) are 8.3\% versus 21.5\%. The dominance of sub-5-minute visits is striking. Possible explanations include brief reconnaissance visits by potential clients assessing the establishment, management check-ins, or quick drop-offs. The deficit in long visits contrasts with legitimate therapeutic massage, where 60- and 90-minute sessions are standard service offerings \citep{field2016massage}. The medium-duration bucket (5--60 min) is also slightly elevated for high-risk establishments, consistent with shorter commercial transactions than those characterizing therapeutic services.

\begin{table}[htbp]
\centering
\caption{Dwell Time by Risk Category}
\label{tab:dwell}
\renewcommand{\arraystretch}{1.2}
\begin{tabular}{lcc}
\hline
\textbf{Duration} & \textbf{High-Risk} & \textbf{Low-Risk} \\
\hline
Short ($<$5 min) & 75.8\% & 65.9\% \\
Medium (5--60 min) & 15.9\% & 12.5\% \\
Long ($>$60 min) & 8.3\% & 21.5\% \\
\hline
\end{tabular}
\end{table}

\subsubsection{Signature 4: Local Clientele}\label{subsec:sig4}

\begin{figure}[htbp]
\centering
\includegraphics[width=0.6\textwidth]{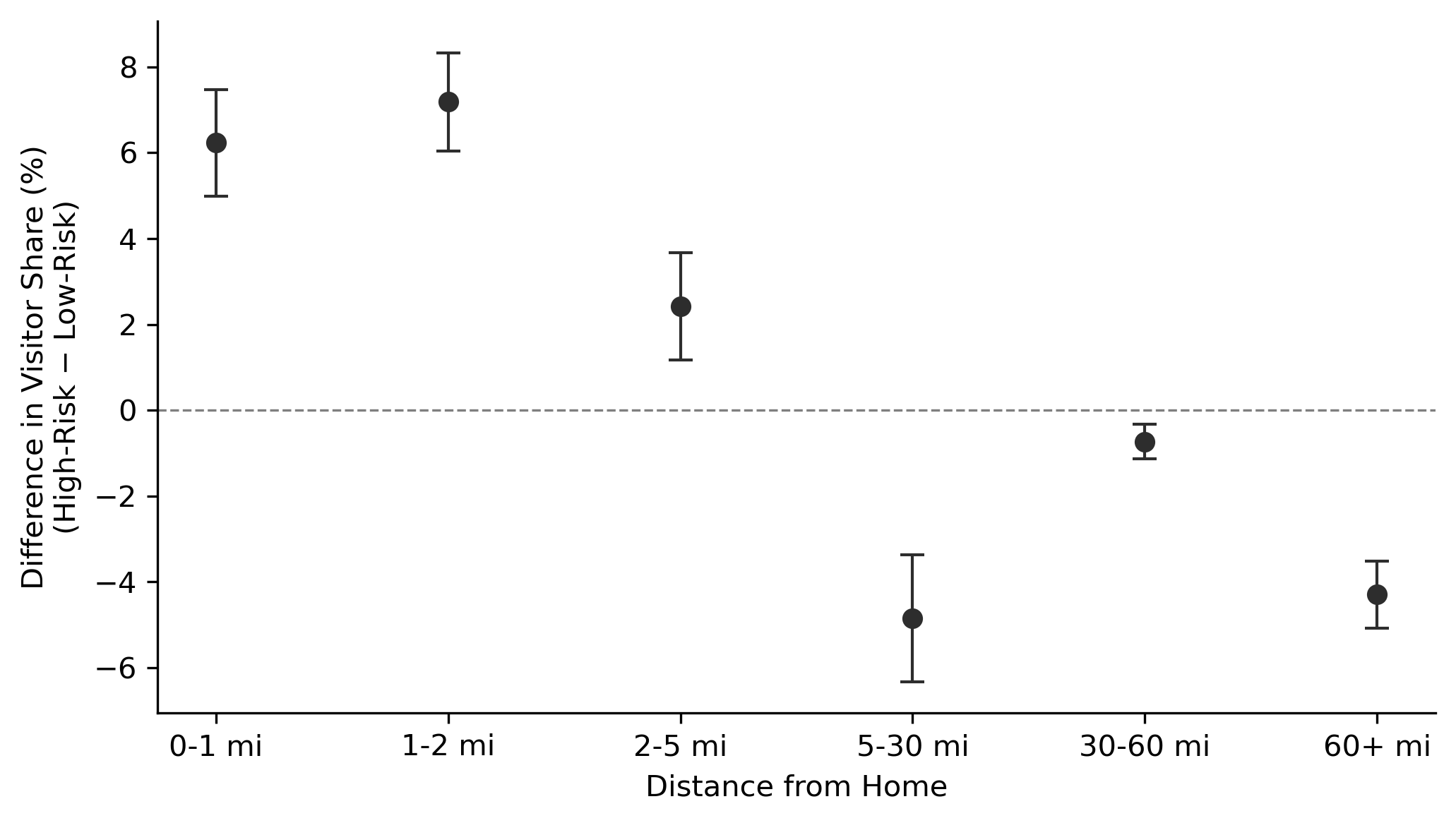}
\caption{Visitor Distance Difference by Risk Category\label{fig:distance}}
\end{figure}

Figure~\ref{fig:distance} shows that high-risk establishments draw disproportionately from nearby areas. The difference exceeds 6 pp within 1 mile and 7 pp for 1--2 miles, reversing at greater distances ($\sim$5 pp lower for 5--30 miles). This geographic concentration reflects the constraints of client acquisition for illicit services. Because illicit operations cannot advertise openly through mainstream channels, client acquisition relies on informal networks and local reputation, producing a geographically concentrated clientele. Legitimate spas, by contrast, may draw from broader areas through online marketing, hotel partnerships, or corporate wellness programs.

\subsubsection{Role of Geographic Context}

Location context features account for 34.8\% of model importance yet show no meaningful difference between high-risk and low-risk observations, suggesting they serve as calibrators rather than direct discriminators. In other words, the same mobility pattern may signal differently in a suburban residential area versus an urban entertainment district.

\begin{table}[htbp]
\centering
\caption{Feature Importance by Category}
\label{tab:category-importance}
\renewcommand{\arraystretch}{1.2}
\begin{tabular}{lcl}
\hline
\textbf{Category} & \textbf{Importance} & \textbf{Pattern in High-Risk} \\
\hline
Operational consistency & 36.4\% & Lower variability \\
Location context & 34.8\% & No difference \\
Temporal patterns$^\dagger$ & 13.3\% & More evening, fewer morning, more weekend \\
Service duration & 9.4\% & More short, fewer long \\
Visit distribution & 2.4\% & Higher entropy \\
Volume control & 2.0\% & No difference \\
Market reach & 1.6\% & More local visitors \\
\hline
\multicolumn{3}{l}{\footnotesize $^\dagger$ Decomposes into time-of-day (10.6\%) and day-of-week (2.7\%).} \\
\end{tabular}
\end{table}

Table~\ref{tab:category-importance} summarizes feature importance by category. Operational consistency and location context together account for over 70\% of importance, though only the former shows directional differences between risk groups.

\subsection{RQ3: Inspection Efficiency}\label{subsec:RQ3}

We apply the trained model to produce establishment-level risk rankings for the full population of massage businesses. This section evaluates aggregation strategies using 5-fold cross-validation at the business level, then quantifies the efficiency gains from model-guided inspection.

\subsubsection{Aggregation Strategy Selection.}

Each business appears across multiple weeks. We compare three strategies for converting week-level scores to a single establishment-level ranking:

\begin{itemize}
\item \textbf{Max}: Peak weekly score. Any suspicious week triggers investigation.
\item \textbf{Mean}: Average score across observed weeks.
\item \textbf{Min}: Minimum weekly score. Requires suspicion throughout.
\end{itemize}

Table~\ref{tab:aggregation-comparison} compares coverage achieved by each method using 5-fold cross-validation at the business level. In each fold, approximately 170 holdout illicit establishments are ranked against 16,000 unlabeled businesses. AUC measures discriminative ability: the probability that a randomly selected illicit business scores higher than a randomly selected unlabeled business. Coverage denotes the percentage of known illicit establishments appearing in the top-ranked set.

\begin{table}[htbp]
  \centering
  \caption{Business-Level Performance by Aggregation Method}
  \label{tab:aggregation-comparison}
  \renewcommand{\arraystretch}{1.2}
  \begin{small}
  \begin{tabular}{lccccc}
  \hline
  \textbf{Aggregation} & \textbf{AUC} & \textbf{Top 1\%} & \textbf{Top 5\%} & \textbf{Top 10\%} & \textbf{Top 20\%} \\
  \hline
  Max  & 0.826 $\pm$ 0.028 & 17.0\% $\pm$ 2.3\% & 40.5\% $\pm$ 4.9\% & 52.8\% $\pm$ 6.7\% & 68.2\% $\pm$ 6.7\% \\
  Mean & 0.701 $\pm$ 0.028 &  7.1\% $\pm$ 2.1\% & 20.5\% $\pm$ 1.9\% & 31.6\% $\pm$ 2.3\% & 46.5\% $\pm$ 4.3\% \\
  Min  & 0.527 $\pm$ 0.010 &  1.4\% $\pm$ 0.6\% &  6.6\% $\pm$ 1.8\% & 12.3\% $\pm$ 1.8\% & 22.1\% $\pm$ 1.6\% \\
  \hline
  \end{tabular}
  \end{small}
\end{table}

\noindent The results show that max aggregation achieves the highest AUC (0.826) and coverage at all budget levels. Illicit signatures manifest intermittently, so preserving the peak weekly signal outperforms averaging across periods of varying signal strength. Model-guided inspection of the top 10\% of ranked establishments surfaces 52.8\% (95\% CI: 47--59\%) of known illicit operations, a 5.3-fold improvement over random inspection. Equivalently, reaching 50\% coverage requires inspecting only 10\% of establishments with model guidance versus 50\% at random, an 80\% reduction in workload. These values are conservative lower bounds, as hidden illicit establishments in the unlabeled pool will also concentrate in top ranks.

\subsubsection{Geographic Distribution.}

\begin{figure}[htbp]
\centering
\includegraphics[width=0.85\textwidth]{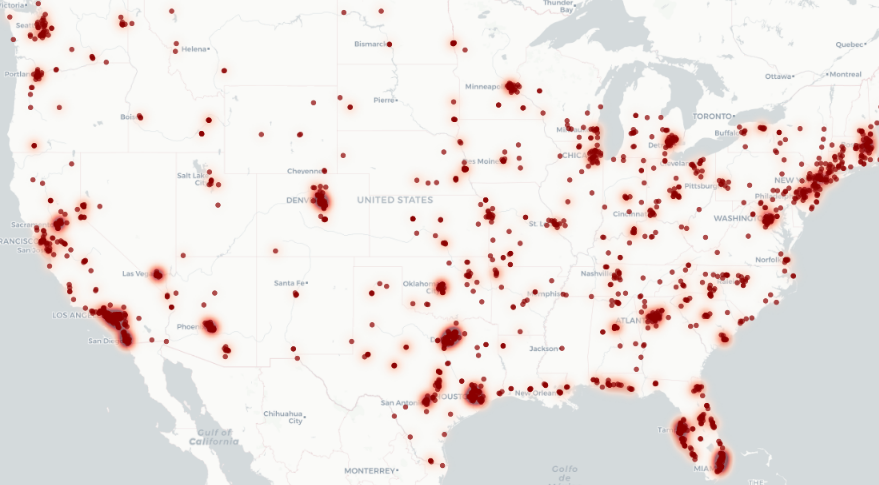}
\caption{Geographic Distribution of High-Risk Establishments\label{fig:risk-map}}
\end{figure}

Figure~\ref{fig:risk-map} displays the spatial distribution of establishments ranked in the top decile by risk score. High-risk establishments concentrate in major metropolitan areas, consistent with prior research on IMB geography \citep{vries_2022_cd, chin2023and}. Dense clusters appear in Los Angeles, the San Francisco Bay Area, Houston, Dallas-Fort Worth, Atlanta, South Florida, and the New York-New Jersey corridor. Clusters appear in major metro areas and along interstate corridors (e.g., the Northeast megalopolis, Florida's I-95 corridor, California's coast), with cross-jurisdictional patterns suggesting potential benefits from coordinated multi-agency task forces. A detailed regional breakdown appears in Appendix~\ref{ec:geographic}.

\section{Discussion and Conclusion}\label{sec:discussion}

This paper develops a mobility-based detection system for identifying illicit massage businesses. We integrate mobility data with positive-unlabeled learning to accommodate the label asymmetry inherent in regulatory enforcement (Section~\ref{subsec:label_asymmetry}).

Our investigation addressed three research questions. For RQ1, mobility data effectively identifies illicit operations, achieving 0.97 AUC and recovering 86.5\% of hidden illicit observations at a threshold of 0.5. For RQ2, four operational signatures distinguish illicit from legitimate establishments (Section~\ref{subsec:rq2}). For RQ3, mobility-based targeting provides over 5-fold efficiency gains, with the top 10\% of ranked establishments capturing 53\% of known illicit operations.

Our work contributes to the literature on regulatory screening under uncertainty. The core methodological insight is that mobility data, unlike online reviews and advertisements, reflects customer behavior that operators cannot curate without fundamentally altering their business operations. This data innovation extends beyond IMB detection to regulatory contexts where illicit facilities hide among legitimate establishments.

The PU learning framework generalizes beyond IMB detection. In any regulatory setting where violations are discovered through investigation but compliance is never directly observed, standard supervised classification introduces systematic bias. Our formulation applies to fraud detection, disease surveillance, and other domains where confirmed labels exist only for positive cases.

For law enforcement agencies, our framework provides a decision-support system that produces calibrated prioritization scores. Rather than responding reactively to complaints or tips, agencies can proactively target high-risk establishments within budget constraints. The ranked output integrates directly with the resource allocation problem in Equation~\eqref{eq:objective}, enabling optimization across jurisdictions.

The four operational signatures have direct interpretive value. Demand stability may indicate that illicit establishments cultivate repeat customers rather than attracting new clients through marketing. Evening concentration may reflect clientele seeking services outside standard business hours. Compressed service duration is consistent with transaction-oriented visits rather than extended therapeutic sessions. Local clientele may indicate word-of-mouth referral networks rather than broad advertising. These patterns appear inherent to the illicit business model and may be more difficult to manipulate than online signals, though this resistance has not been empirically tested. Alternative explanations merit consideration: these patterns may characterize ASW-advertising establishments specifically rather than all illicit operations, and some legitimate massage businesses serving evening clientele or local customers may exhibit similar signatures.

Several limitations warrant consideration. First, our ground truth assumes that advertising on adult services websites constitutes evidence of illicit activity. While advertising on these platforms provides strong evidence of commercial sex services, no labeling heuristic is without error. Some advertisements could be posted by third parties without the establishment's knowledge, and the platform's self-selection dynamics mean our positive labels may not represent the full diversity of illicit operations. Second, our temporal coverage spans January through December 2024, a single calendar year. This temporal scope introduces several considerations. Mobility patterns may exhibit seasonal variation that our cross-sectional analysis does not capture; establishments may show different visitation patterns during holiday seasons or summer months. Patterns may also shift as enforcement practices evolve and operators adapt their strategies over time. Future research could examine temporal stability by training on earlier periods and evaluating on later periods to assess model robustness to distributional drift. Third, we cannot distinguish human trafficking from voluntary commercial sex work; our framework identifies establishments offering commercial sex services, not the coercive conditions under which workers operate. This distinction matters for enforcement priorities and ethical considerations. Fourth, mobility data raises privacy concerns. While aggregated and anonymized, the data reflects individual movement patterns. We use the data only for aggregate establishment-level analysis and do not attempt to identify individual visitors. Deployment would require appropriate institutional oversight and privacy safeguards.

Several extensions are worth considering. First, real-time monitoring can enable dynamic enforcement that responds to changing operational patterns. Extension to other illicit business types (unlicensed cannabis dispensaries, illegal gambling operations, unlicensed medical clinics) would test the generalizability of mobility-based detection. Second, causal analysis of enforcement interventions could inform optimal targeting strategies that account for displacement effects and network disruption. Finally, integration with other data sources (licensing records, complaint histories, financial transaction patterns) may improve detection beyond what mobility data alone provides.

\section*{Data Availability Statement}
The mobility data used in this study are available through a commercial subscription to Advan Research via the Dewey Data platform (\url{https://www.deweydata.io/}). The adult services website data contain sensitive information related to ongoing law enforcement investigations and cannot be publicly shared. Researchers interested in replicating or extending this work should contact the authors to discuss access arrangements.

\section*{Acknowledgment}
This work is partially supported by the National Science Foundation.

\bibliographystyle{plainnat}
\bibliography{imb-mobility-clean}

\appendix
\counterwithin{figure}{section}
\counterwithin{table}{section}

\section{Merged Data Structure}\label{ec:merged_data}

\begin{figure}[htbp]
\centering
\includegraphics[width=1\textwidth]{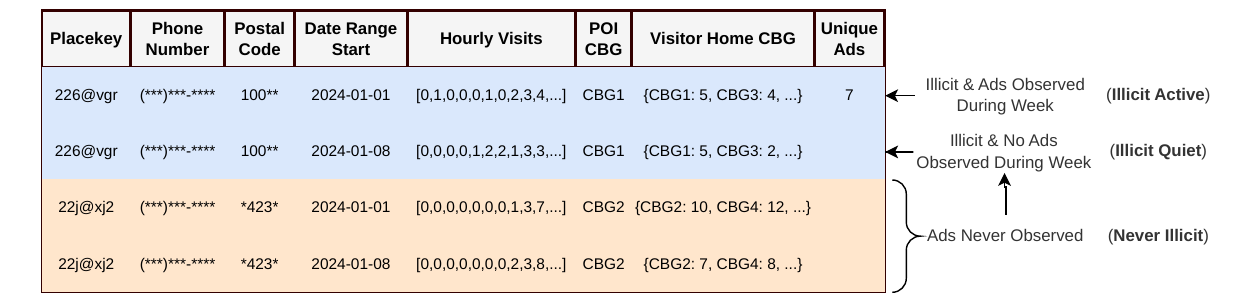}
\caption{Merged Data Example\label{fig:ec-merged-data}}
\end{figure}

Figure~\ref{fig:ec-merged-data} provides a fictional example of a subset of the merged dataset to illustrate important structural characteristics and our labeling convention. The table contains eight columns representing a subset of the fields available in the merged dataset. The \textit{Placekey} column provides a standardized location identifier for each POI. \textit{Phone Number} and \textit{Postal Code} are shown masked for privacy. \textit{Date Range Start} specifies the beginning of each weekly observation period. \textit{Hourly Visits} contains an array of 168 values representing visitor counts for each hour of the week. \textit{POI CBG} identifies the Census Block Group where the establishment is located, while \textit{Visitor Home CBG} is a dictionary mapping visitor home CBGs to visit counts. For example, \{CBG1: 5, CBG3: 4, ...\} indicates that five visitors to this establishment during the week resided in CBG1 and four resided in CBG3. \textit{Unique Ads} contains the count of distinct advertisements observed for that location during the corresponding week, when applicable.

The rows are color-coded to indicate two categories of business: 1) those linked to ASW ads (blue, the top two rows) and 2) those not linked to ASW ads (orange, the bottom two rows). Among the businesses with observed ads, weeks with positive values in the \textit{Unique Ads} column correspond to \textit{Illicit Active} entries, indicating that ads were observed during those weeks. Among the businesses with observed ads, weeks with missing values in the \textit{Unique Ads} column correspond to \textit{Illicit Quiet} entries, indicating that there is evidence of illicit activity at the location but no advertising activity during that week. Orange rows represent locations where advertisements were never observed (\textit{Never-ASW}). Although these establishments cannot be linked to an ASW ad, that alone is not sufficient to conclude that they are non-illicit, i.e., the absence of evidence is not evidence of absence.

\section{Why Standard Classification Fails}\label{ec:standard_fails}

A naive approach treats $\mathcal{P}$ as positives and $\mathcal{U}$ as negatives, then trains a standard classifier. This introduces systematic bias. Let $f: \mathbb{R}^d \rightarrow [0,1]$ be a classifier trained under this assumption. The expected loss becomes:

\begin{equation}
    \mathbb{E}[\ell(f)] = \mathbb{E}_{x \sim \mathcal{P}}[\ell(f(x), 1)] + \mathbb{E}_{x \sim \mathcal{U}}[\ell(f(x), 0)]
\end{equation}

\noindent The second term is problematic: we penalize the classifier for assigning high scores to instances in $\mathcal{U}$, but some of these are true positives. The classifier learns to distinguish \textit{platform-advertised} establishments from others, rather than \textit{illicit} from \textit{legitimate}. Any hidden illicit establishment in $\mathcal{U}$ that resembles those in $\mathcal{P}$ will persist in the negative class and push the decision boundary away from the true discrimination.

\section{Geographic Patterns in Risk Distribution}\label{ec:geographic}

Beyond individual metropolitan areas, the risk map (Figure~\ref{fig:risk-map} in the main text) reveals broader regional patterns. California shows heavy concentration throughout the state, extending along the coast from San Diego through the Bay Area. Texas exhibits multi-city clustering, with high-risk establishments distributed across Houston, Dallas-Fort Worth, San Antonio, and Austin. The Eastern Seaboard displays a near-continuous corridor of activity from Boston through Washington, D.C., reflecting the interconnected nature of the Northeast megalopolis. Florida's pattern follows the I-95 corridor, with particular density in the Miami metropolitan area extending north through Orlando to Jacksonville.

The distribution also highlights geographic variation in apparent risk. Mountain West states (Montana, Wyoming, the Dakotas) show markedly sparse activity, while Sunbelt states exhibit higher overall concentration. Secondary clusters in Phoenix, Seattle, Denver, Las Vegas, Chicago, and Detroit indicate that the phenomenon extends beyond the largest markets. These patterns carry enforcement implications: clusters crossing state lines (such as the New York-New Jersey corridor or the Dallas-Fort Worth metroplex spanning multiple counties) may benefit from coordinated multi-jurisdictional task forces rather than isolated local efforts.

\section{Detailed Feature Definitions}\label{app:features}

This appendix provides complete specifications for the 28 features engineered from mobility data. All features are computed at the POI-week level unless otherwise noted.

\subsection{Temporal Visit Patterns (8 Features)}

We partition the 168-hour week into six non-overlapping 4-hour windows and compute the proportion of visits in each:

\begin{table}[h]
\centering
\caption{Temporal Window Definitions}
\label{tab:ec-temporal-features}
\begin{tabular}{lll}
\hline
\textbf{Feature Name} & \textbf{Hours (Local Time)} & \textbf{Computation} \\
\hline
\texttt{early\_morning} & 04:00--08:00 & $\sum_{h \in [4,8)} v_h / \sum_h v_h$ \\
\texttt{morning\_business} & 08:00--12:00 & $\sum_{h \in [8,12)} v_h / \sum_h v_h$ \\
\texttt{afternoon} & 12:00--16:00 & $\sum_{h \in [12,16)} v_h / \sum_h v_h$ \\
\texttt{evening} & 16:00--20:00 & $\sum_{h \in [16,20)} v_h / \sum_h v_h$ \\
\texttt{late\_evening} & 20:00--24:00 & $\sum_{h \in [20,24)} v_h / \sum_h v_h$ \\
\texttt{late\_night} & 00:00--04:00 & $\sum_{h \in [0,4)} v_h / \sum_h v_h$ \\
\hline
\end{tabular}
\end{table}

\noindent where $v_h$ denotes total visits during hour $h$ across all seven days of the week.

Two additional temporal features capture day-of-week patterns:
\begin{itemize}
    \item \texttt{weekend}: Proportion of visits occurring on Saturday or Sunday
    \item \texttt{friday\_sat}: Proportion of visits occurring on Friday or Saturday
\end{itemize}

\subsection{Visit Distribution Features (3 Features)}

\textbf{Hourly entropy.} We aggregate visits across all seven days to obtain a 24-hour distribution. Let $p_h$ denote the proportion of weekly visits in hour $h \in \{0, 1, \ldots, 23\}$:
\begin{equation}
    H_{\text{hourly}} = -\sum_{h=0}^{23} p_h \log(p_h)
\end{equation}
Higher values indicate visits spread evenly across hours; lower values indicate concentration in fewer hours. Maximum entropy occurs when visits are uniform across all 24 hours.

\textbf{Daily entropy.} We aggregate across hours to obtain a seven-day distribution. Let $q_d$ denote the proportion of weekly visits on day $d \in \{0, 1, \ldots, 6\}$:
\begin{equation}
    H_{\text{daily}} = -\sum_{d=0}^{6} q_d \log(q_d)
\end{equation}

\textbf{Peak hour ratio.} For each hour of the day, we sum visits across all seven days. The peak hour ratio is:
\begin{equation}
    \text{Peak ratio} = \frac{\max_h \sum_{d} v_{h,d}}{\sum_{h,d} v_{h,d}}
\end{equation}
High values indicate a pronounced daily rush period.

\subsection{Service Duration Features (3 Features)}

The Advan data provides bucketed dwell time distributions. We compute proportions across three categories:
\begin{itemize}
    \item \texttt{short\_visit\_share}: Proportion of visits with dwell time $<5$ minutes
    \item \texttt{medium\_visit\_share}: Proportion of visits with dwell time 5--60 minutes
    \item \texttt{long\_visit\_share}: Proportion of visits with dwell time $>60$ minutes
\end{itemize}

\subsection{Market Reach Features (6 Features)}

For each visitor, we compute the great-circle distance from their home CBG centroid to the POI using the Haversine formula:
\begin{equation}
    d = 2r \arcsin\left(\sqrt{\sin^2\left(\frac{\phi_2 - \phi_1}{2}\right) + \cos(\phi_1)\cos(\phi_2)\sin^2\left(\frac{\lambda_2 - \lambda_1}{2}\right)}\right)
\end{equation}
where $r = 3{,}958.756$ miles is Earth's radius, and $(\phi, \lambda)$ denote latitude and longitude in radians.

We bin distances into six categories and compute the proportion of visitors in each:

\begin{table}[h]
\centering
\caption{Visitor Distance Categories}
\label{tab:ec-distance-features}
\begin{tabular}{lll}
\hline
\textbf{Feature Name} & \textbf{Distance Range} & \textbf{Market Interpretation} \\
\hline
\texttt{dist\_0\_1mi} & 0--1 miles & Very local / walkable \\
\texttt{dist\_1\_2mi} & 1--2 miles & Neighborhood \\
\texttt{dist\_2\_5mi} & 2--5 miles & Local trade area \\
\texttt{dist\_5\_30mi} & 5--30 miles & Regional \\
\texttt{dist\_30\_60mi} & 30--60 miles & Extended regional \\
\texttt{dist\_60plus} & $>$60 miles & Out-of-area \\
\hline
\end{tabular}
\end{table}

\subsection{Operational Consistency Features (5 Features)}

These features are computed at the POI level across all observed weeks and remain constant for all weekly observations of a given establishment. Normalization by mean weekly visits makes these metrics scale-invariant, enabling comparison across establishments with different visitor volumes. Let $V = \{v_1, v_2, \ldots, v_T\}$ denote weekly visit counts for an establishment observed over $T$ weeks.

\textbf{Coefficient of variation:}
\begin{equation}
    \text{CV} = \frac{\sigma^{V}}{\mu^{V}}.
\end{equation}
Higher values indicate more volatile, less predictable traffic.

\textbf{Trend:} Slope of weekly visits over time, normalized by the mean:
\begin{equation}
    \text{Trend} = \frac{\beta_1}{\mu^{V}}
\end{equation}
where $\beta_1$ is the OLS slope from regressing $v_t$ on $t$. Positive values indicate growing traffic over the observation period; negative values indicate declining operations.

\textbf{Burstiness:} Standard deviation of week-over-week changes, normalized:
\begin{equation}
    \text{Burstiness} = \frac{\sigma^{\Delta V}}{\mu^{V}}
\end{equation}
where $\Delta V = \{v_2 - v_1, v_3 - v_2, \ldots\}$. Lower values indicate stable week-to-week demand; higher values indicate irregular traffic with sudden surges or drops.

\textbf{Maximum jump:} Largest single-week change, normalized:
\begin{equation}
    \text{Max jump} = \frac{\max_t |v_{t+1} - v_t|}{\text{mean}(V)}
\end{equation}
High values indicate susceptibility to demand shocks; lower values suggest resilient, predictable operations.

\textbf{Active ratio:} Fraction of weeks with any recorded visits:
\begin{equation}
    \text{Active ratio} = \frac{|\{t : v_t > 0\}|}{T}
\end{equation}
Values close to 1.0 indicate continuous operation throughout the observation period; lower values indicate intermittent activity with periods of apparent closure.

\subsection{Location Context Features (2 Features)}

\textbf{CBG area:} Log-transformed land area (square miles) of the establishment's census block group:
\begin{equation}
    \text{Log CBG area} = \log(\text{ALAND} / 2{,}589{,}988)
\end{equation}
where ALAND is the land area in square meters from Census TIGER files.

\textbf{Partisan index:} County-level partisanship measure from the National Neighborhood Data Archive \citep{ICPSR38506}, calculated by averaging presidential and Senate vote ratios over four consecutive election cycles. Values range from 0 (strongly Democratic) to 1 (strongly Republican).

\subsection{Volume Control (1 Feature)}

\textbf{Log weekly visits:} Natural logarithm of total visits during the week:
\begin{equation}
    \text{Log visits} = \log(1 + \sum_h v_h)
\end{equation}
The $+1$ adjustment handles weeks with zero recorded visits.

\end{document}